\documentclass{article}
\usepackage{spconf,amsmath,graphicx}
\usepackage{multirow}
\usepackage{amsmath}
\usepackage{amssymb}
\usepackage{svg}
\usepackage{tikz}
\usepackage{soul}

\title{Deep Multiway Canonical Correlation Analysis for \\ Multi-subject  EEG Normalization}
\name{Jaswanth Reddy Katthi, Sriram Ganapathy\thanks{This work started at the Telluride Neuromorphic Cognition Engineering Workshop $2019$ held during the summer of $2019$ sponsored by the NSF. This work was performed with grants from the   Department of Science and Technology, India grants under the Early Career Research Award (ECR01341) and  the  Department of Atomic Energy project 34/20/12/2018-BRNS/34088 }}
\address{
 Learning and Extraction of Acoustic Patterns (LEAP) lab, Indian Institute of Science, Bangalore.\\}
 
\begin{document}
\ninept
\maketitle
\begin{abstract}
The normalization of brain recordings from multiple subjects responding to the natural stimuli is one of the key challenges in auditory neuroscience. The objective of this normalization is to transform the brain data in such a way as to remove the inter-subject redundancies and to boost the  component related to the stimuli.  In this paper, we propose a deep learning framework to improve the correlation of electroencephalography (EEG) data recorded from multiple subjects engaged in an audio listening task. The proposed model extends the linear multi-way canonical correlation analysis (CCA)  for audio-EEG analysis using an auto-encoder network with a shared  encoder layer. The model is trained to optimize a combined loss involving correlation and reconstruction. The experiments are performed on EEG data collected from subjects listening to natural speech and music. In these experiments, we show that the proposed deep multi-way CCA (DMCCA) based model significantly improves the correlations over the linear multi-way CCA approach with absolute improvements of $0.08$ and  $0.29$ in terms of the Pearson correlation values for speech and music tasks respectively.     
\end{abstract}
\section{Introduction}
\label{sec:intro}
The analysis of brain signals as a response to an  auditory stimulus is of profound interest from both the understanding as well as the application perspective. The main goal of this research is to find a relationship between the stimulus and the response.  
The response signal  in non-invasive measurements is typically in the form of electroencephalography (EEG), magnetoencephalography (MEG) or, functional magnetic resonance imaging (fMRI). Among these modalities, the EEG represents the most easily accessible brain recordings for analyzing speech comprehension and audio signal perception \cite{wostmann2017tracking}.    

Though the EEG can provide high temporal resolution (typically sampled at $128$ Hz or more),  the data is captured at the scalp. As a result, the desired response is significantly attenuated by the skull and makes the recordings highly noisy  (average SNR of $-20$dB~\cite{sanei2007eeg}). For short stimuli, it is a common practice to repeat the experiment for a given condition. The averaging operation on the brain responses improves the SNR. This resultant average response is termed as the event related potential (ERP)~\cite{coles1995event}.  However, for naturalistic auditory stimuli like speech and audio, repetition of the experiments with the same conditions is often impractical. This has brought in greater demand for single-trial EEG analysis methods. 

The temporal response function (TRF)~\cite{lalor2009resolving} is one of the earliest in this direction of single-trial EEG analysis. This method assumes a linear relationship between the stimuli and the EEG response. The system identification is performed using least-squares and the  performance of these models is quantified by the Pearson correlation between the estimated signal and the actual signal. In the EEG recordings, the presence of signals unrelated to the presented stimuli, as well as the attenuation by the scalp, results in shallow correlation values for these models in the range of $0.1-0.2$~\cite{lalor2009resolving}. 

The canonical correlation analysis (CCA) tries to alleviate this problem to some extent by finding common transformations of the stimuli and the  responses. It finds a pair of optimum linear transforms, for the stimulus and the response, such that the correlation between the two new projections is maximized~\cite{thompson1984canonical}. 
In order to characterize the non-linear relationship between the stimuli and EEG response, we had recently proposed a deep CCA~\cite{andrew2013deep} for audio-EEG datasets~\cite{katthi2020deep}.

Most of the single-trial analysis methods model the stimulus-response data only for a single subject. One way to expand the dataset is to perform the experiment on multiple subjects for the same stimuli and conditions. At the same time, as many other subject-specific factors drive the EEG response, the problem of finding common signals across the subjects' responses is challenging. The linear CCA can be performed only on two views of the data at a time. A multi-variate version of the CCA, termed as multiway CCA  or generalized CCA ~\cite{correa2010multi,fu2017scalable,zhang2017inter}, has been proposed for aggregating EEG response from multiple-views (subjects).   As all views (EEG responses) are based on the same stimulus, the components that  are common across the views~\cite{parra2018multi} can elicit the information regarding the brain processing of the stimuli.  The multiway CCA (MCCA)  successfully extracts common signals across more than two data-views ~\cite{de2019multiway}. However, all these models (TRF, CCA and MCCA) are based on linear assumptions of the brain functions. In this paper, we hypothesize that a linear system-response assumption may be too simplistic and inadequate to uncover the transfer function of the brain responses. 

In this paper, we propose a deep learning model for the multi-subject EEG normalization task. Given the EEG response to the same stimuli from multiple subjects, the proposed model projects the EEG responses to a space that maximizes the inter-subject correlation. This model, termed as deep multiway-CCA, extends the linear methods to allow the analysis of single-trial multi-subject audio-EEG datasets. The model uses a reconstruction based regularization to  alleviate the issues of over-fitting in the deep MCCA model. The deep MCCA model is compared with linear MCCA~\cite{de2019multiway,parra2018multi} for speech~\cite{di2015low} and audio~\cite{gang2017decoding} perception tasks. 
In our experiments comparing the linear MCCA with the proposed DMCCA, the DMCCA model improves correlation on the average by $0.08$ for speech tasks and $0.18$ for audio tasks. These improvements in the correlation values are found to be statistically significant ($p<0.01$) on a pairwise t-test analysis as well. 

\section{Related Prior Work}
\label{sec:prior}
The recent years have seen the use of deep learning for several brain mapping tasks like computational memory prediction \cite{sun2016remembered} and for reconstruction of brain activity for visual stimuli \cite{yuan2019wave2vec}. A review of several efforts in decoding brain activity using deep learning techniques is given in Zheng et al. \cite{zheng2020decoding}. 

In auditory tasks, EEG recordings have shown to contain rhythm information in music perception~\cite{stober2014using}. Recently, Das et al.~\cite{das2020linear} showed that deep learning techniques can improve auditory attention decoding in the perception of noisy speech.  In multi-speaker cocktail party scenarios, Deckers et al.~\cite{deckers2018eeg} showed that neural networks can identify the attended speaker.  
  
In the context of CCA, the extensions to connect multiple data matrices have been proposed under names such as multiple CCA, multiway CCA, multiset CCA, generalized CCA, or multiway orthonormalized partial least squares~\cite{gross2015collaborative,sturm2016analyzing,zhang2011multiway}. There have been attempts to generalize the deep CCA~\cite{benton2017deep}. Liu et al.~\cite{liu2020efficient} has proposed a deep model, referred to as DGCCA~\cite{benton2017deep} in the literature, for SSVEP frequency recognition. To the best of our knowledge, this paper presents the first efforts in developing deep models for normalization of audio-EEG data from multiple subjects. 

\section{Deep MCCA}
\label{sec:mathback}

\subsection{Background - Linear Multi-way CCA}
\label{subsec:MCCA}
The multiway CCA (MCCA) is a linear method that expands the linear CCA to more than two data-views. It obtains an optimum linear transform for each data-view, such that the inter-set correlation (ISC) across all the projections is maximum~\cite{parra2018multi, de2019multiway}.  

Let $N$ data-views be denoted as $\mathbf{x}_n \in \mathcal{R}^{d_n} $ for $n = 1 \text{ to } N$ and $D_N=\sum_{n}d_n$. The MCCA tries to extract the signals common across the data-views. For $1$D projection, the MCCA projects the $N$ data-views onto a $1$D subspace, such that all the views are highly correlated to each other. Let,  the vector $\mathbf{v}_n \in \mathcal{R}^{d_n \times 1}$ denote projection vector that linearly transforms data-view $\mathbf{x}_n$ onto the $1$D subspace. The linear transforms are learnt such that the inter-set correlation (ISC) among the projections is maximum. The ISC can be formulated as: 
\begin{equation}
\text{ISC} =\frac{1}{N-1} \frac{r_{B}}{r_{W}}
\end{equation}
where $r_B$ and $r_W$ are the between-set covariance and within-set covariance respectively given by,
\begin{eqnarray}
r_B & = & \sum_{i=1}^{N} \sum_{j=1, j \neq i}^{N} \mathbf{v}_i^\top \mathbf{R}^{i j} \mathbf{v}_{j} \\
r_W & = & \sum_{i=1}^{N} \mathbf{v}_i^\top \mathbf{R}^{i i} \mathbf{v}_{i}.
\end{eqnarray}
Here, $\mathbf{R}^{i j} \in \mathcal{R}^{d_i \times d_j}$ are the cross-correlation matrix of the data-views $\mathbf{x}_i$ and $\mathbf{x}_j$.

The optimal linear transform vectors $\{\mathbf{v}_n\}_{n=1}^{N}$ are obtained by finding the eigenvector with the maximum eigenvalue of the following eigen problem~\cite{parra2018multi}: 
\begin{equation}
\label{eq:eigen}
\mathbf{R} \mathbf{v} = \lambda \mathbf{D} \mathbf{v}
\end{equation}
where $\lambda=(N-1) \rho+1$, $\mathbf{v} \in \mathcal{R}^{D_N \times 1}$ is the concatenated vector of the linear transform vectors $\{\mathbf{v}_n\}_{n=1}^{N}$, matrix $\mathbf{R} \in \mathcal{R}^{D_N \times D_N}$ is a block matrix with its block elements as $[\mathbf{R}]_{ij} = \mathbf{R}^{ij}$  and $\mathbf{D} \in \mathcal{R}^{D_N \times D_N}$ is a block-diagonal matrix with its elements as $[\mathbf{D}]_{ii} = \mathbf{R}^{ii}$. If the $N$ data-views are to be projected onto a $d$ dimensional ($d>1$) subspace, then the linear transform matrix for each data-view $\mathbf{x}_n$ is $\mathbf{V}_n \in \mathcal{R}^{d_n \times d}$, obtained by choosing the top `$d$' eigenvectors of Eq.~\ref{eq:eigen} . 



\subsection{Deep Multiway CCA}
\label{subsec:DMCCA}


\subsubsection{Natural Extension - DGCCA}
\begin{figure}[t!]
  \includegraphics[width=8.6cm]{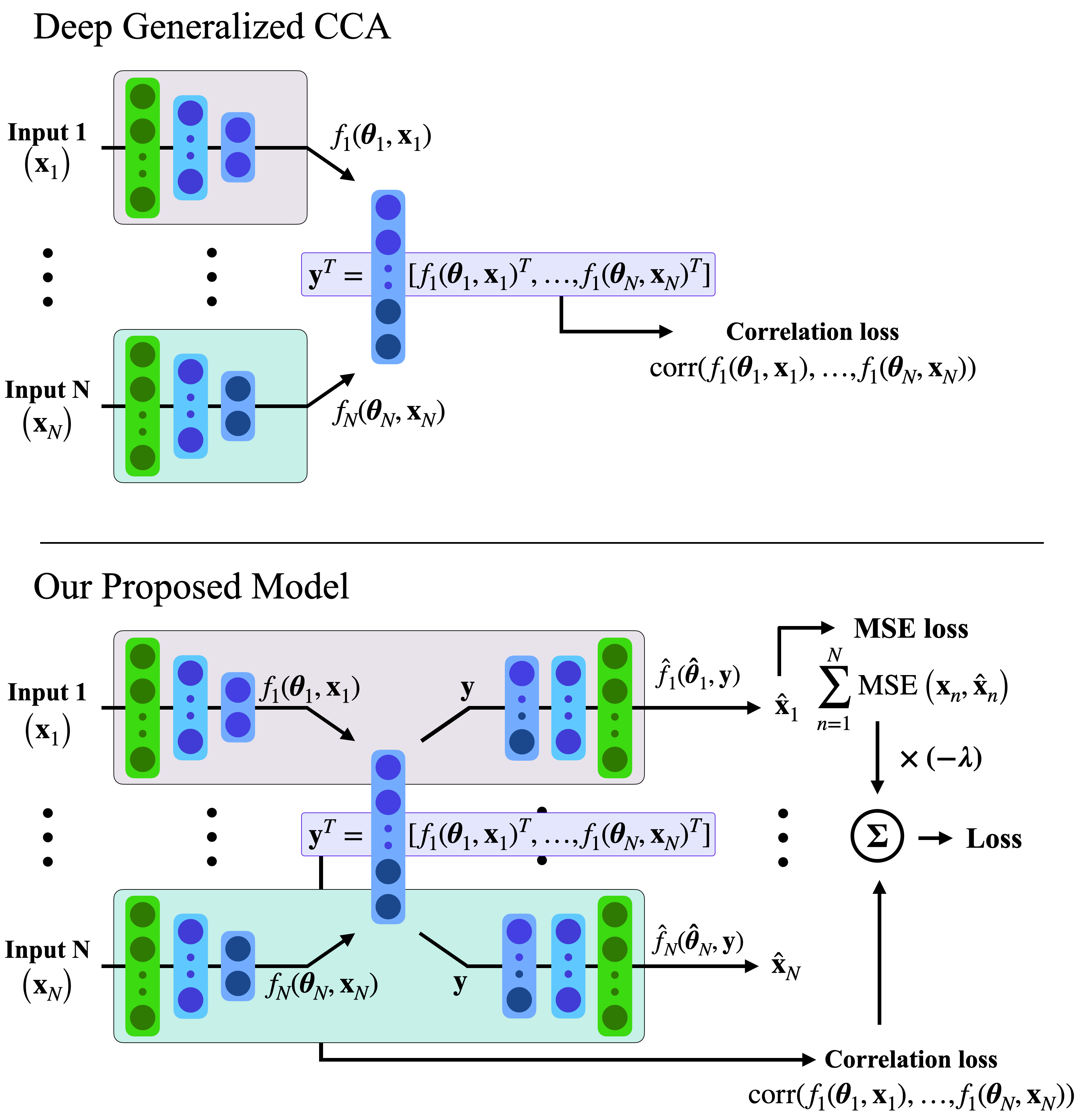}
  \centering
  \caption{The DGCCA and the proposed DMCCA.}
    \label{fig:dmcca_model}
\end{figure}

Using the $N$ random multi-variates,  the goal of DGCCA is to determine $N$ neural transformations (non-linear) such that the correlation cost is maximized. Let $f_n(\cdot)$, for $n = 1 \text{ to } N$,  denote the neural network that takes $\mathbf{x}_n$ as input and outputs a $d$ dimension representation. Let each  neural network's trainable parameters be denoted as  $\boldsymbol{\theta}_n$.
The parameters are learned such that the sum of all inter-variate correlations is maximized. This correlation cost is,
\begin{align}
    \rho = \sum_{i=1}^{N} \sum_{j=1, j \neq i}^{N} \operatorname{corr}\left(f_{i}\left(\mathbf{x}_i ; \boldsymbol{\theta}_i\right), f_{j}\left(\mathbf{x}_j ; \boldsymbol{\theta}_j\right)\right)
    \label{rho}
\end{align}
We refer to this model as the deep generalized canonical correlation analysis (DGCCA) model in Figure~\ref{fig:dmcca_model}.  


\subsubsection{Proposed Model}
The $N$ neural networks in the DGCCA model transform the $d_n$ dimensional data-views to $d$ ($d\leq d_n$) dimensional common subspace.  This transformation can be viewed as a shared encoder of the views. In the proposed model, we also construct a decoder which attempts to reconstruct the data-views from the shared encoder output. This architecture is also illustrated in Figure~\ref{fig:dmcca_model}.

The encoder model consists of $N$ neural transformations, each operating on  one of the $N$ data-views. The $N$ data-views, propagated through the encoding parts,  generate $N$ vectors of $d$ dimensions ($f_{i}\left(\mathbf{x}_i ; \boldsymbol{\theta}_i\right)$) which are concatenated to form a $Nd$ dimensional vector (shared encoder output). This $Nd$ dimensional encoder output ($\mathbf{y}$) is used as the common representation for all the decoder layers. The decoder also has $N$ parallel networks, each of which attempts to reconstruct the corresponding data-view $\mathbf{x}_n$. This proposed model is referred to as the deep multiway canonical correlation analysis (DMCCA) model. 
 
The DMCCA model is trained to maximize the inter-set correlation among the encoders' outputs, and minimize the MSE (Mean Square Error) loss at each of decoder outputs. Let the trainable parameters of the model be split into $N$ parts, where each $\{\boldsymbol{\theta}_n\,, \boldsymbol{\hat{\theta}}_n\}_{n=1}^{N}$ corresponds to the parameters of the encoder and decoder of the $n^{\text{th}}$ data-view. Then, the  optimization of the DMCCA model is performed by maximizing
\begin{align}
    \rho' = \rho - \lambda \sum_{n=1}^{N}\operatorname{MSE}\left(\mathbf{x}_n, \hat{f}_n(\mathbf{y} ; \boldsymbol{\hat{\theta}}_n)\right)
\end{align}
where $\rho$ is defined by Eq.~\eqref{rho} and $\operatorname{MSE}(\cdot)$ denotes the mean squared error between the reconstructed data-view and the input data-view (auto-encoding loss). The parameter $\lambda$ is a hyper-parameter that controls the relative importance of the two-cost functions. It is found that the MSE regularization loss is complementary to the inter-variate correlation loss. Further, the DGCCA is a special case of DMCCA for $\lambda=0$.

Following the model training, the encoder part of each data-view  is used as the fixed non-linear transform that normalizes each data-view to remove subject-specific redundancies in the data and boost the stimuli related components.

\section{Audio-EEG Setup}
\label{sec:audio-eeg}
\subsection{Datasets}
\label{subsec:datasets}
We perform experiments on two naturalistic auditory stimuli and their respective EEG responses: a speech dataset and an audio dataset. The speech dataset is taken from a study on cortical entrainment to natural speech~\cite{di2015low}. The audio dataset is the NMED-H which consists of EEG responses from $48$ subjects listening to natural music~\cite{gang2017decoding}. 

The speech dataset is the same dataset used for the baseline linear MCCA model~\cite{de2019multiway}. This dataset is also used in the application of CCA~\cite{de2018decoding} and deep CCA~\cite{katthi2020deep} models. The EEG data are recorded using $128$ electrodes. We perform the same pre-processing steps done by Cheveigné et al.~\cite{de2019multiway}.  The EEG data are down-sampled to $64$ Hz and de-trended and de-noised using noise tools software~\cite{de2018robust}. Then, EEG data are passed through a band-pass filter with a pass-band in the range of $0.1-12$ Hz. The stimuli data are obtained from audio sampled at $44100$ Hz. The audio envelope is squared and smoothed by convolving the envelope with a square window~\cite{de2018robust}. Finally, the stimuli data (audio envelopes) are downsampled to $64$ Hz and are further compressed with a power of $1/3$.  More details about the pre-processing can be found in Cheveigné et al.~\cite{de2018decoding}.   

The audio dataset is the NMED-H dataset~\cite{gang2017decoding}. It is designed to study the brain signals for a natural music listening task. It contains EEG responses to intact and scrambled versions of $4$ full-length Hindi pop songs. Each stimulus is approximately $4.5$ minutes long, sampled at $44.1k$Hz. The EEG data are recorded using $125$ electrodes. They are sampled at $125$ Hz with average reference. More details of the  data acquisition and pre-processing are given in Kaneshiro~\cite{kaneshiro2016toward}. The features from the audio stimuli are extracted in a similar way as described by Gang et al.~\cite{gang2017decoding}. The acoustic features are extracted using the music information retrieval (MIR) toolbox~\cite{lartillot2007matlab}. 
We have extracted $20$ dimensional features in $25$ms analysis windows with a $50$\% overlap between frames~\cite{tzanetakis2002musical}. 
The Principal component analysis (PCA) is performed on these $20$D data to extract a $1$D representation (PC$1$) for the data. Along with the PC$1$, two individual features, root mean square (RMS) and spectral flux, that reflect amplitude envelope and timbre respectively are chosen to obtain a $3$D representation for the stimuli. The final sampling rate of the acoustic features is $80$ Hz. Similarly, the EEG response is resampled to $80$ Hz. All experiments on the audio dataset are performed on each of these $3$D pre-processed features or the $1$D envelope features.

\subsection{Performance metric}
\label{subsec:models}

The outputs of linear/deep MCCA models are provided to a linear CCA based model. It is applied to further transform the MCCA/DMCCA model outputs and the stimulus data, separately for each subject. We use the CCA3 configuration described in Cheveigné et al.~\cite{de2018decoding}.

The performance comparison between the linear and deep versions of the MCCA models is done using the final correlation scores of the first canonical covariate of the CCA3 method. Another metric to measure the models' performance is the classification accuracy. For that, we have performed a Cohen's d-prime match-vs-mismatch classification analysis as discussed by Cheveigné et al.~\cite{de2018decoding}.

\subsection{DMCCA Model}
\label{subsec:dmcca_eeg}
For EEG response from $N$ subjects and a $d_s$ time-lagged $1$D (common) stimulus, the encoder section consists of two hidden layers with $60$ nodes each and an output layer of $10$ nodes. Thus, the final de-noised representations for the EEG responses  is $10$D. The $N+1$ ($N$ EEG responses plus the stimuli features) vectors of $10$D are concatenated to form a $10(N+1)$D vector. This vector is provided to each decoder. The decoder section also consists of two hidden layers of $60$ nodes and $110$ nodes respectively. 



The final outputs from the linear MCCA are of $128$D and $125$D for speech and audio data respectively (same as the input). The outputs from the deep MCCA are of $10$D. The linear MCCA first encodes the data onto $d_s$ dimensions and then projects them onto the $128$D space~\cite{de2019multiway} (for speech dataset) or $125$D {\color{blue} ( } for the audio dataset {\color{blue} ) }. The choice of time-lag ($d_s$) for each version of MCCA is done by varying it from $10$ to $110$. The time-lag which provided the best performance is chosen.

To make the final $1$D representations of both the MCCA versions comparable, the $10$D denoised EEG response of each subject is provided to the filterbank from the CCA$3$ method~\cite{de2018decoding} and a PCA to project them onto a $139$D subspace. The $10$D stimulus features are projected onto a $1$D subspace using PCA and provided to the filterbank, resulting in a $21$D stimulus features. These features are provided to a linear CCA model to obtain $1$D representations~\cite{de2018decoding}.


\subsection{Experimental Setup}
\label{subsec:setup}

\begin{figure}[t!]
  \includegraphics[width=8.8cm]{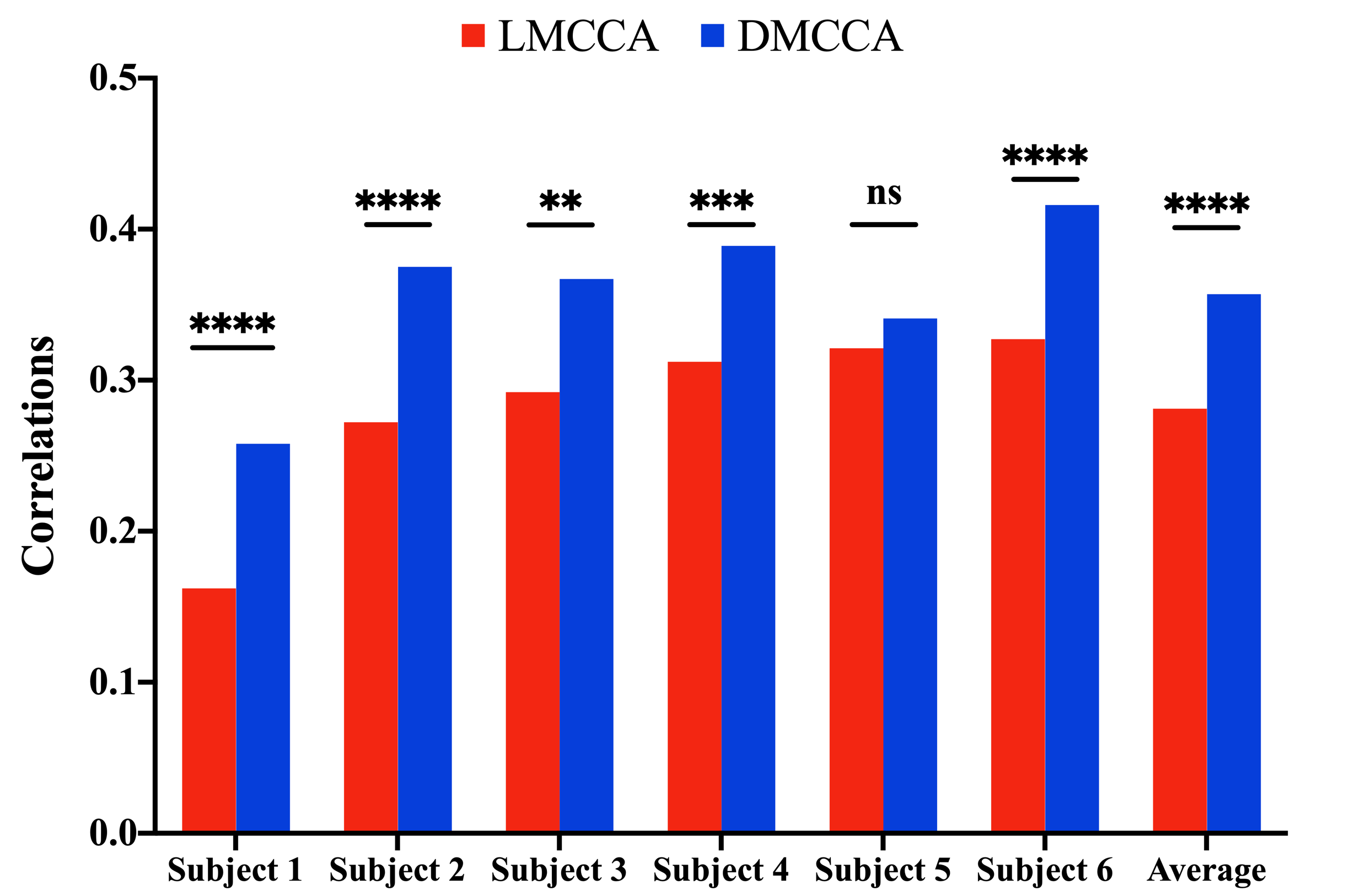}
  \centering
  \caption{The correlation values for linear and deep MCCA, on $6$ subjects from the speech dataset. Pairwise t-test statistical significance is shown (ns implies no significance ($p>0.01$), ** implies $p\leq0.01$), *** implies $p\leq0.001$), **** implies $p\leq1e-4$) }
    \label{fig:lmcca_vs_dmcca}
\end{figure}

From the speech dataset, stimulus-response of $6$ subjects is used in the experiments. Each subject has $20$ recorded sessions with approximately $160$ seconds in each session. We perform $20$ fold-validation experiments on these $20$ sessions. In each validation experiment, $18$ of the $20$ sessions are used for training the models, while one of the remaining sessions is used for validation and the other for testing. This results in about $200k$ samples for training the model.

The audio dataset has EEG response from $48$ subjects listening to $4$ versions of $4$ full length Hindi pop songs, each having about $4.5$ minutes duration. 
We aggregated the stimulus-response data based on the common stimuli. Each subject has $2$ EEG recording trials for $2$ different stimuli. 
This results in about $38k$ samples for training and $2k$ samples for validation and testing respectively.

The $N$ subjects EEG responses and a time-lagged stimulus (time lag of $40$ for linear and $60$ for deep MCCA) features are provided to the linear and deep versions of MCCA. The denoised outputs of the MCCA (linear/deep) are provided to the CCA3 method~\cite{de2018decoding}.
The final representations of $1$D are obtained by the CCA$3$ method and their correlations are used for comparing the MCCA methods.

A leaky ReLU activation function with a negative slope coefficient of $0.1$ is used at the encoder layer in DMCCA model. A linear activation function is used at the output layer of the decoder sections. The dropout regularization~\cite{srivastava2014dropout} is incorporated in the deep models, and the dropout with the best perfomance is chosen. 

The MCCA (linear/deep) denoising step is performed on the combination of EEG responses whereas the final step of CCA3 method is done for each subject separately. 


\section{Results}
\label{sec:results}

\subsection{Performance analysis}
For speech dataset, the linear and deep MCCA methods are performed on a collection of $6$ subjects. The results of the cross-validation experiments for all the $6$ subjects are presented in Fig. ~\ref{fig:lmcca_vs_dmcca}. As seen in the Fig.~\ref{fig:lmcca_vs_dmcca}, the audio-EEG correlation is improved for all the subjects using DMCCA. On the average, the DMCCA improves the correlation by around $0.08$. 
Additional experiments comparing the performance of the DGCCA model~\cite{benton2017deep}, proposed by Liu et al.~\cite{liu2020efficient} and our proposed DMCCA model has been done on the speech dataset. The average correlation of the the six subjects has increased from $0.158$ (for the DGCCA model) to $0.234$ (our proposed model). This shows that this MSE regularization term helps the model improve the  correlations.

For the audio dataset, the $48$ subjects were divided into sets based on the common stimulus, with $12$ subjects in each set. The MCCA methods were tried on these sets separately. One experiment with the preprocessed envelope representing the stimuli features and $3$ experiments with the $3$D features (from ~\ref{subsec:datasets}) were performed for both versions of MCCA. The results are presented in the Table.~\ref{table:nmedh}. The DMCCA consistently improves over linear MCCA, on an average of $0.29$ in the correlation values.

\begin{table}[t!]
\caption{Average correlation values of $48$ subjects from NMED-H Dataset}
\label{table:nmedh}
\begin{center}
\begin{tabular}{ |c|c|c| }
    \hline
     & LMCCA & DMCCA \\
    \hline
    Envelope      & 0.0694 & 0.3268 \\
    PC1           & 0.0237 & 0.3552 \\
    RMS           & 0.0305 & 0.3169 \\
    Spectral Flux & 0.0432 & 0.3261  \\
    \hline
\end{tabular}
\vspace{-0.5cm}
\end{center}
\end{table}


\begin{table}[t!]
\caption{The d-prime metric for both the datasets (in percentages)}
\label{table:d_prime}
\begin{center}
\begin{tabular}{ |c||c|c||c|c| }
    \hline
     & \multicolumn{2}{c||}{Speech dataset} & \multicolumn{2}{c|}{NMED-H} \\
    \hline
    segment size & LMCCA & DMCCA & LMCCA & DMCCA \\
    \hline
    1s  &  2.84 &  2.91 &  1.36 &   2.12 \\
    2s  &  3.02 &  4.27 &  3.45 &   4.80 \\
    4s  &  5.39 &  7.52 &  7.18 &  11.84 \\
    8s  &  8.45 & 11.29 & 14.94 &  34.18 \\
    16s & 15.87 & 18.57 & 32.38 &  38.39 \\
    32s & 19.12 & 32.90 &   -   &   -    \\
    \hline
\end{tabular}
\vspace{-0.5cm}
\end{center}
\end{table}

\subsection{Statistical Analysis}
\label{subsec:stats}
To test the statistical significance, one tail pairwise t-test is performed on the final correlation values for the linear and deep versions of MCCA. For the speech dataset, it is seen that the improvements are statistically significant ($p<1e-2$) for $5$ out of $6$ subjects.
It is found to be statistically significant ($p<1e-2)$ for all the $4$ stimulus features for the audio dataset. Similar to Cheveigné et al.~\cite{de2018decoding}, Cohen's d-prime value for match-vs-mismatch classification analysis is also performed. Segments of EEG and stimuli of duration $T$ are selected and the correlation among all the combinations of EEG and stimuli are measured. The correlation of an EEG response and its related stimulus falls under the 'match' class, whereas for unrelated stimuli fall under 'mismatch' class. For $T$ varying from $1$ to $32$ seconds, the d-prime values for the MCCA methods, for both the datasets, are tabulated in Table.~\ref{table:d_prime}. As seen in Table. ~\ref{table:d_prime}, for all the segment durations, the deep MCCA improves over the linear version for both the datasets.

\section{Summary}
In this paper, we have proposed a deep multi-way CCA framework for multi-subject EEG analysis of auditory stimuli. The DMCCA model uses the EEG data from multiple subjects in an autoencoder setup with a shared layer. The model uses a combination of reconstruction loss and correlation loss. The experiments on the speech dataset shows that the model significantly improves the correlations over the linear multi-way CCA. A classification based measure of stimuli-response alignment also shows the benefits of the proposed DMCCA model. Further, we show that the improvements seen in speech data are consistent for EEG-music datasets as well. 

%




\bibliographystyle{IEEEbib}
\bibliography{strings,refs}

\begin{thebibliography}{10}

\bibitem{wostmann2017tracking}
Malte W{\"o}stmann, Lorenz Fiedler, and Jonas Obleser,
\newblock ``Tracking the signal, cracking the code: Speech and speech
  comprehension in non-invasive human electrophysiology,''
\newblock {\em Language, Cognition and Neuroscience}, vol. 32, no. 7, pp.
  855--869, 2017.

\bibitem{sanei2007eeg}
Saeid Sanei and Jonathon~A Chambers,
\newblock ``{EEG} signal processing,''
\newblock {\em Wiley Online Library}, 2007.

\bibitem{coles1995event}
Michael~GH Coles and Michael~D Rugg,
\newblock {\em Event-related brain potentials: An introduction.},
\newblock Oxford University Press, 1995.

\bibitem{lalor2009resolving}
Edmund~C Lalor, Alan~J Power, Richard~B Reilly, and John~J Foxe,
\newblock ``Resolving precise temporal processing properties of the auditory
  system using continuous stimuli,''
\newblock {\em Journal of neurophysiology}, vol. 102, no. 1, pp. 349--359,
  2009.

\bibitem{thompson1984canonical}
Bruce Thompson,
\newblock {\em Canonical correlation analysis: Uses and interpretation},
\newblock Number~47. Sage, 1984.

\bibitem{andrew2013deep}
Galen Andrew, Raman Arora, Jeff Bilmes, and Karen Livescu,
\newblock ``Deep canonical correlation analysis,''
\newblock in {\em International Conference on Machine Learning}, 2013, pp.
  1247--1255.

\bibitem{katthi2020deep}
Jaswanth~Reddy Katthi, Sriram Ganapathy, Sandeep Kothinti, and Malcolm Slaney,
\newblock ``Deep canonical correlation analysis for decoding the auditory
  brain,''
\newblock in {\em 2020 42nd Annual International Conference of the IEEE
  Engineering in Medicine \& Biology Society (EMBC)}. IEEE, 2020, pp.
  3505--3508.

\bibitem{correa2010multi}
Nicolle~M Correa, Tom Eichele, T{\"u}lay Adal{\i}, Yi-Ou Li, and Vince~D
  Calhoun,
\newblock ``Multi-set canonical correlation analysis for the fusion of
  concurrent single trial erp and functional mri,''
\newblock {\em Neuroimage}, vol. 50, no. 4, pp. 1438--1445, 2010.

\bibitem{fu2017scalable}
Xiao Fu, Kejun Huang, Mingyi Hong, Nicholas~D Sidiropoulos, and Anthony Man-Cho
  So,
\newblock ``Scalable and flexible multiview max-var canonical correlation
  analysis,''
\newblock {\em IEEE Transactions on Signal Processing}, vol. 65, no. 16, pp.
  4150--4165, 2017.

\bibitem{zhang2017inter}
Qiong Zhang, Jelmer~P Borst, Robert~E Kass, and John~R Anderson,
\newblock ``Inter-subject alignment of meg datasets in a common
  representational space,''
\newblock {\em Human brain mapping}, vol. 38, no. 9, pp. 4287--4301, 2017.

\bibitem{parra2018multi}
Lucas~C Parra,
\newblock ``Multi-set canonical correlation analysis simply explained,''
\newblock {\em arXiv preprint arXiv:1802.03759}, 2018.

\bibitem{de2019multiway}
Alain de~Cheveign{\'e}, Giovanni~M Di~Liberto, Doroth{\'e}e Arzounian,
  Daniel~DE Wong, Jens Hjortkj{\ae}r, S{\o}ren Fuglsang, and Lucas~C Parra,
\newblock ``Multiway canonical correlation analysis of brain data,''
\newblock {\em NeuroImage}, vol. 186, pp. 728--740, 2019.

\bibitem{di2015low}
Giovanni~M Di~Liberto, James~A O’Sullivan, and Edmund~C Lalor,
\newblock ``Low-frequency cortical entrainment to speech reflects phoneme-level
  processing,''
\newblock {\em Current Biology}, vol. 25, no. 19, pp. 2457--2465, 2015.

\bibitem{gang2017decoding}
Nick Gang, Blair Kaneshiro, Jonathan Berger, and Jacek~P Dmochowski,
\newblock ``Decoding neurally relevant musical features using canonical
  correlation analysis.,''
\newblock in {\em ISMIR}, 2017, pp. 131--138.

\bibitem{sun2016remembered}
Xuyun Sun, Cunle Qian, Zhongqin Chen, Zhaohui Wu, Benyan Luo, and Gang Pan,
\newblock ``Remembered or forgotten?—an eeg-based computational prediction
  approach,''
\newblock {\em PloS one}, vol. 11, no. 12, pp. e0167497, 2016.

\bibitem{yuan2019wave2vec}
Ye~Yuan, Guangxu Xun, Qiuling Suo, Kebin Jia, and Aidong Zhang,
\newblock ``Wave2vec: Deep representation learning for clinical temporal
  data,''
\newblock {\em Neurocomputing}, vol. 324, pp. 31--42, 2019.

\bibitem{zheng2020decoding}
Xiao Zheng, Wanzhong Chen, Mingyang Li, Tao Zhang, Yang You, and Yun Jiang,
\newblock ``Decoding human brain activity with deep learning,''
\newblock {\em Biomedical Signal Processing and Control}, vol. 56, pp. 101730,
  2020.

\bibitem{stober2014using}
Sebastian Stober, Daniel~J Cameron, and Jessica~A Grahn,
\newblock ``Using convolutional neural networks to recognize rhythm stimuli
  from electroencephalography recordings,''
\newblock in {\em Advances in neural information processing systems}, 2014, pp.
  1449--1457.

\bibitem{das2020linear}
Neetha Das, Jeroen Zegers, Tom Francart, Alexander Bertrand, et~al.,
\newblock ``Linear versus deep learning methods for noisy speech separation for
  eeg-informed attention decoding,''
\newblock {\em Journal of Neural Engineering}, vol. 17, no. 4, pp. 046039,
  2020.

\bibitem{deckers2018eeg}
Lucas Deckers, Neetha Das, Amir~Hossein Ansari, Alexander Bertrand, and Tom
  Francart,
\newblock ``Eeg-based detection of the attended speaker and the locus of
  auditory attention with convolutional neural networks,''
\newblock {\em bioRxiv}, p. 475673, 2018.

\bibitem{gross2015collaborative}
Samuel~M Gross and Robert Tibshirani,
\newblock ``Collaborative regression,''
\newblock {\em Biostatistics}, vol. 16, no. 2, pp. 326--338, 2015.

\bibitem{sturm2016analyzing}
Irene Sturm,
\newblock {\em Analyzing the perception of natural music with EEG and ECoG},
\newblock Ph.D. thesis, Berlin, Technische Universit{\"a}t Berlin, 2016.

\bibitem{zhang2011multiway}
Yu~Zhang, Guoxu Zhou, Qibin Zhao, Akinari Onishi, Jing Jin, Xingyu Wang, and
  Andrzej Cichocki,
\newblock ``Multiway canonical correlation analysis for frequency components
  recognition in ssvep-based bcis,''
\newblock in {\em International conference on neural information processing}.
  Springer, 2011, pp. 287--295.

\bibitem{benton2017deep}
Adrian Benton, Huda Khayrallah, Biman Gujral, Dee~Ann Reisinger, Sheng Zhang,
  and Raman Arora,
\newblock ``Deep generalized canonical correlation analysis,''
\newblock {\em arXiv preprint arXiv:1702.02519}, 2017.

\bibitem{liu2020efficient}
Qianqian Liu, Yong Jiao, Yangyang Miao, Cili Zuo, Xingyu Wang, Andrzej
  Cichocki, and Jing Jin,
\newblock ``Efficient representations of eeg signals for ssvep frequency
  recognition based on deep multiset cca,''
\newblock {\em Neurocomputing}, vol. 378, pp. 36--44, 2020.

\bibitem{de2018decoding}
Alain de~Cheveign{\'e}, Daniel~DE Wong, Giovanni~M Di~Liberto, Jens Hjortkjaer,
  Malcolm Slaney, and Edmund Lalor,
\newblock ``Decoding the auditory brain with canonical component analysis,''
\newblock {\em NeuroImage}, vol. 172, pp. 206--216, 2018.

\bibitem{de2018robust}
Alain de~Cheveign{\'e} and Doroth{\'e}e Arzounian,
\newblock ``Robust detrending, rereferencing, outlier detection, and inpainting
  for multichannel data,''
\newblock {\em NeuroImage}, vol. 172, pp. 903--912, 2018.

\bibitem{kaneshiro2016toward}
Blair~Bohannan Kaneshiro,
\newblock {\em Toward an Objective Neurophysiological Measure of Musical
  Engagement},
\newblock Ph.D. thesis, Stanford University, 2016.

\bibitem{lartillot2007matlab}
Olivier Lartillot and Petri Toiviainen,
\newblock ``A matlab toolbox for musical feature extraction from audio,''
\newblock in {\em International conference on digital audio effects}. Bordeaux,
  2007, pp. 237--244.

\bibitem{tzanetakis2002musical}
George Tzanetakis and Perry Cook,
\newblock ``Musical genre classification of audio signals,''
\newblock {\em IEEE Transactions on speech and audio processing}, vol. 10, no.
  5, pp. 293--302, 2002.

\bibitem{srivastava2014dropout}
Nitish Srivastava, Geoffrey Hinton, Alex Krizhevsky, Ilya Sutskever, and Ruslan
  Salakhutdinov,
\newblock ``Dropout: a simple way to prevent neural networks from
  overfitting,''
\newblock {\em The journal of machine learning research}, vol. 15, no. 1, pp.
  1929--1958, 2014.

\end{thebibliography}

\end{document}